 \documentstyle[prl,aps]{revtex}  
\def\bra{\langle} \def\ket{\rangle} \def\ack{\,|\,}

\begin{document}
\draft


 \twocolumn[\hsize\textwidth\columnwidth\hsize  
 \csname @twocolumnfalse\endcsname              

\title{
Microscopic Description of Band Structure at Very Extended 
Shapes in the $A \sim 110$ Mass Region 
} 
\author{Ching-Tsai Lee$^1$, Yang Sun$^1$, Jing-ye Zhang$^1$,
Mike Guidry$^1$, Cheng-Li Wu$^2$
}
\address{
$^1$Department of Physics and Astronomy, University of Tennessee,
Knoxville, Tennessee 37996 \\
$^2$National Center for Theoretical Science,
Hsinchu, Taiwan 300, ROC 
}

\date{\today}
\maketitle

\begin{abstract}
Recent experiments have 
confirmed the existence of rotational bands in the $A \sim 110$ mass region 
with very extended shapes lying between super- and hyper-deformation.
Using the projected shell model, we make a first attempt to describe
quantitatively such a band structure in $^{108}$Cd. 
Excellent agreement is achieved in the
dynamic moment of inertia $\Im^{(2)}$ calculation. 
This allows us to suggest the spin
values for the energy levels, which are experimentally unknown. 
It is found that, at this
large deformation, 
the sharply down-sloping orbitals in the proton $i_{13/2}$ subshell 
are responsible for the irregularity in the  
experimental $\Im^{(2)}$, and the wave functions of the observed states 
have a dominant component of  
two-quasiparticles from these orbitals. 
Measurement of transition quadrupole moments 
and g-factors will test these findings, and thus
can provide a deeper understanding 
of the band structure at very extended shapes.
\end{abstract}


 ]  

\narrowtext

Low-energy nuclear physics is characterized by a rich 
shell structure. 
A nuclear system 
can be stabilized at large deformations by quantum shell effects, 
leading to many observable phenomena. 
Nuclear super- and hyper-deformations have long been predicted to exist 
in nuclei. 
The study of superdeformed nuclei has been an active focus of 
low-energy nuclear physics for the past two decades. 
Today, superdeformation (SD) at high spin is
not an isolated phenomenon, but instead is observed
across the nuclear periodic table \cite{HW99}. 
However, the field has remained active because of the continuing 
discovery of new, exotic bands with distinct characteristic structures that one 
has never encountered before. 
 
The very recent experimental work of Clark {\it et al.} \cite{Exp} 
in the $A\sim 110$ mass region provides an example.   
Early calculations \cite{NPF76,WD95,Ch01} suggested 
the likely regions where super- and hyper-deformation might exist.
In particular, calculations predicted a very extended shape minimum 
in $^{108}$Cd, lying 
intermediate
between super- and hyper-deformation, and that the nuclei 
at this deformation are possibly stable against fission. In a recent 
letter \cite{Exp}, 
Clark {\it et al.} have found a rotational band 
in $^{108}$Cd with the data 
indicating that the nucleus has the most deformed structure
identified to date.  
 
An irregular pattern in dynamic moments of inertia ($\Im^{(2)}$)
was observed in this highly deformed band \cite{Exp}.  
The experimental $\Im^{(2)}$ curve begins with a large value 
$(\approx 80 [\hbar^2 \mbox{MeV}^{-1}])$ at lower spins 
and quickly drops to 50.  
For the deduced transition quadrupole moment $Q_t$, the lower 
experimental limit 
was given as 9.5 $e$b \cite{Exp}. 
Besides these global features, no detailed 
structure information is currently known for this band.   
This Rapid Communication aims to analyze the microscopic structure 
of this band and to explain the cause of the rapid
change in $\Im^{(2)}$. Our calculations suggest that 
the low-K $i_{13/2}$ proton orbitals are responsible for the $\Im^{(2)}$ 
irregularity
and that the two-quasiparticle configurations from these orbitals 
dominate the structure of the observed states. 
Based on the excellent reproduction of the experimental 
$\Im^{(2)}$, we further suggest 
the spin values for the observed energy levels.  
 
Our analysis is performed using the projected shell model (PSM) \cite{HS95}.
The PSM follows closely the shell model philosophy, and in fact,
is a spherical shell model truncated in a deformed
BCS single-particle basis. The truncation is first
achieved within the quasi-particle basis with respect
to the deformed BCS vacuum
$|0 \rangle$; then rotational
symmetry that is violated in the deformed mean-field basis
is restored by
standard angular-momentum projection techniques
\cite{RS80} to
form a spherical basis in the laboratory frame; finally the shell model
Hamiltonian is diagonalized in the projected basis. The
truncation obtained in this way is very efficient, 
making a shell model calculation for a heavy, very deformed nucleus possible.
Moreover,    
spin is a strictly conserved quantity in a theory with
angular momentum projection. This is the most desired feature when
applying a model to study SD structure because, for the most observed 
SD bands, it
is very difficult 
to determine spin values experimentally. 
In the past, the PSM was consistently 
applied for SD structure study 
to the mass regions $A\sim 190$ \cite{Sun97,Hg194-g},    
$A\sim 130$ \cite{Sun95}, $A\sim 80$ \cite{Ted}, $A\sim 60$ \cite{Sun99}, 
and $A\sim 30$ \cite{Sun01}. 
The predicted spin values for some SD bands were confirmed by later
experiments. 

In the PSM, 
the many-body wave function is a superposition of
(angular momentum) projected multi-quasiparticle states,
\begin{equation}
| \psi^I_M \rangle ~=~
\sum_{\kappa} f_{\kappa} \hat P^I_{MK_\kappa}
| \varphi_{\kappa} \rangle ,
\label{ansatz}
\end{equation}
where $\hat P^I_{MK}$ is the projection operator \cite{RS80} and 
$| \varphi_{\kappa} \rangle$ denotes basis states consisting of
the quasiparticle (qp) vacuum, 2 quasi-neutron and -proton, and
4-qp states 
\begin{eqnarray}
\ack\varphi_\kappa\rangle=\left\{ \ack 0 \rangle,
\ a^\dagger_{\nu_1}a^\dagger_{\nu_2}\ack 0\rangle,
\ a^\dagger_{\pi_1}a^\dagger_{\pi_2}\ack 0\rangle,
\ a^\dagger_{\nu_1}a^\dagger_{\nu_2}a^\dagger_{\pi_1}
a^\dagger_{\pi_2}\ack 0\rangle \right\}.
\label{conf}
\end{eqnarray}
In Eq. (\ref{conf}), 
$a^\dagger$'s are the qp creation operators, $\nu$'s ($\pi$'s)
denote the neutron (proton) Nilsson quantum numbers, which run over
properly selected (low-lying) orbitals and $\ack 0 \rangle$ is the
qp vacuum (0-qp state).
The dimensionality of the qp basis
(\ref{conf}) is about 80 and the deformation of the basis is 
fixed at  
$\epsilon_2 = 0.67$ for calculations in this paper.  
The choice of this deformation is based on experimental information
\cite{Exp}. 
We note that the choice of a basis deformation is not very sensitive  
for the PSM because the input deformation parameter serves to provide a good
basis and the real deformation will be determined dynamically 
by later configuration mixing. 

For this $N=60$ and $Z=48$ nucleus with a large deformation, 
single-particle orbitals from several major shells can be found near
the Fermi levels. Among them, the high-$j$ orbitals $i_{13/2}$ and
$h_{11/2}$ are expected to play an important role. 
However, some orbitals from the major shells
$N=4$ and $N=3$ also surround the Fermi levels. For a proper treatment,
we have therefore included four major shells ($N=3, 4, 5$ and 6) 
each for neutrons and protons. This configuration space will ensure
a correct description not only for the moment of inertia, but also for
the quadrupole moment, which receives contributions from 
many nucleons.  
The importance of the neutron $i_{13/2}$ and proton $h_{11/2}$ 
orbitals was mentioned in Ref. \cite{Exp}.
However, we have found that the structure of the measured states
is mainly determined by the $i_{13/2}$ protons, as one will see below.     

The same Hamiltonian (including spherical single-particle, 
residual quadrupole--quadrupole, monopole pairing, and quadrupole
pairing terms) as employed in all of the early PSM work 
is then 
diagonalized in the basis (\ref{ansatz}).  
The strength of the quadrupole-quadrupole force is determined 
such that it has a self-consistent relation with
the quadrupole deformation $\epsilon_2$ \cite{HS95}.
The monopole-pairing force
constants $G_M$ used in the present calculations are
\begin{equation}
\begin{array}{c}
G_{\mbox{\scriptsize M}} =
\left[ 17.10 \mp 11.16 \frac{N-Z}{A}\right] ~A^{-1} ,
\label{GMONO}
\end{array}
\end{equation}
with ``$-$" for neutrons and ``$+$" for
protons. These are the same as those 
in the SD calculations for the $A\sim 80$ mass region \cite{Ted}, 
and have an 8$\%$ reduction from the values used for the 
$A\sim 190$ mass region \cite{Sun97,Hg194-g}.
Finally,
the strength parameter $G_Q$ for the quadrupole pairing is simply taken
to be proportional to $G_M$. For the present calculation a standard  
proportionality constant of 0.16
is used.

We have applied this formalism to calculate energy levels of $^{108}$Cd
for each spin and use the resulting wave functions to compute
the matrix elements. 
In Fig. 1, the calculated energy levels are compared with data \cite{Exp}. 
As in most SD band experiments,
the spin value of
the measured energy levels
was not determined experimentally because of lack of information of linking them
to the known low-lying states.
Assuming $\Im^{(1)} \approx \Im^{(2)}$, the authors in \cite{Exp} estimated
that the spin range for the 11 observed $\gamma$-rays is $I = 40(2)\hbar
- 60(2)\hbar$. The calculated
transition energies E$\gamma$ are displayed in Fig. 1(a)
together with data. 
The best fit between experiment and theory is achieved
if we place the measured 
first E$\gamma$ at $I=36\hbar$. We regard this as fairly
good agreement in spin values
between our calculation and the experimental estimation. 

Static and dynamic moments of inertia, $\Im^{(1)}$ and $\Im^{(2)}$,
are compared in Fig. 1(b).
The most characteristic feature is that the theoretical $\Im^{(2)}$ exhibits
a peak, starting from $I=10\hbar$ and ending at $I=50\hbar$, 
with the centroid around $I=36\hbar$. 
This corresponds to the clear slope change in the $\Im^{(1)}$ curve 
in the same spin range. 
As can be seen, the measured $\Im^{(2)}$ falls nicely onto the right side  
of the peak in the theoretical curve 
if the spin values $I = 36\hbar$ for the measured first E$\gamma$
is assumed.  
With the same assumption, the experimental $\Im^{(1)}$ coincides with
our calculated curve rather precisely. 

The validity of the condition $\Im^{(1)} \approx \Im^{(2)}$ can be
further examined for the whole spin range.
It is seen in Fig. 1(b) that the condition is nicely fulfilled at the 
beginning and at the end of the theoretical curves beyond the peak region. 
However, $\Im^{(1)}$ and $\Im^{(2)}$
differ strongly for a wide range of spin values in between. 
The measured cascade of $\gamma$-rays stops at the top of the peak 
and the experiment did not identify any further states along the left
side of the $\Im^{(2)}$ peak. 

The existence of such a peak 
in $\Im^{(2)}$ usually indicates a crossing of
two (or more) bands with strong band interaction. 
The crossing can change the content of the wave functions for the spin
states before and after the crossing. It is thus of interest to determine  
what the crossing bands are.
To see this, a band diagram \cite{HS95} 
is the most useful tool. 

In the PSM, a band diagram
is defined as the rotational energies of a set of bands calculated as
the expectation values of the Hamiltonian with respect to projected 
quasiparticle states $| \varphi_{\kappa} \rangle$ of Eq. (\ref{conf}). 
These are the PSM basis bands before configuration mixing.    
Fig. 2 shows the band diagram from the current calculation.
We plot the ground (0-qp) band, three important 2-qp bands, and
two representative 4-qp bands. In addition, the lowest (yrast) states for
each spin obtained from band mixing are plotted as dots.
In this diagram, we look for bands that are the lowest in energy (thus most
influential on the yrast states) in the spin range of interest.  
It can be seen that the 
three 2-qp bands cross the ground band and become lower in energy   
at high spins. These bands are built on 
two 2-quasineutron states of the $i_{13/2}$ orbitals,
coupled to ($K=1/2 \otimes K=3/2$) and ($K=3/2 \otimes K=5/2$) 
respectively, and
one 2-quasiproton state of the $i_{13/2}$ orbitals, 
coupled to ($K=1/2 \otimes K=3/2$).   
One sees that the proton 2-qp states (solid line in Fig. 2) 
define the lowest 2-qp band in the spin region 
where the first band crossing occurs.
The inset to Fig. 2 shows clearly that the proton 2-qp band crosses
the ground band at $I=38\hbar$. 
The other two 2-qp bands associated with $i_{13/2}$ neutrons lie roughly 1 MeV
higher than the proton 2-qp band and they cross the ground band at higher
spin around $I=46\hbar$. The 2-qp bands based on $h_{11/2}$ protons (not shown
in Fig. 2) lie even higher, about 2 MeV above the lowest proton $i_{13/2}$ 
2-qp band. Thus, the band diagram tells us that the proton 
$i_{13/2}$ 2-qp states may be the dominant component in the
observed band structure. 
 
As one has seen in Fig. 1, the measured $\gamma$-rays are suggested to be 
in the spin range $I=36\hbar - 56\hbar$. For this spin range, the 
proton 2-qp band is the lowest configuration, as shown in Fig. 2. 
We thus expect that the $i_{13/2}$ protons with $K=1/2$ and 3/2 
should strongly influence the high spin properties of the observed states,
leading to additional observable phenomena.
Electromagnetic transitions are important in understanding the underlying 
physics and in testing theoretical models. 
Transition quadrupole moments measure the contribution of many orbits 
to the collective structure. 
A large polarization effect is normally expected if the wave functions
contain significant components from particular orbitals.  
This should be seen in calculations of the total electrical quadrupole moment. 
On the other hand, the g-factor is
more sensitive to the single-particle structure in wave functions
because of the intrinsically opposite signs of the neutron and proton
$g_s$.
Therefore, we now discuss two additional physical
quantities, the transitional quadrupole moment $Q_t$ and the g-factor. 

The transition quadrupole moment $Q_t$ is related to the $B(E2)$
reduced transition probability through
\begin{equation}
Q_t (I) = \left( \frac{16 \pi}{5} \frac{ B(E2,I\rightarrow I-2)} 
{\bra I K 2 0 | I-2 K \ket} \right)^{1/2} .
\end{equation}
The reduced transition probabilities $B(E2)$ from the initial state
$\psi^{I_i}$ to the final state $\psi^{I_f}$ are given by
\begin{equation}
B(E2,I_i \rightarrow I_f) = {\frac {e_{\rm{eff}}^2} {2 I_i + 1}}
| \bra \psi^{I_f} || \hat Q || \psi^{I_i} \ket |^2 .
\end{equation}
In the calculations, we have used effective charges ($e_{\rm{eff}}$)
of 1.5e for protons
and 0.5e for neutrons, which are the same as in the previous PSM calculations 
\cite{HS95}.

The g-factor can be expressed as   
\begin{equation}
 g(I) = {{\bra\psi^I_{M=I} | \hat\mu_z | \psi^I_{M=I}\ket} \over {\mu_N I}}  
      = {{\bra\psi^I || \hat\mu || \psi^I\ket} \over {\mu_N \sqrt{I(I+1)}}}, 
\end{equation}
with $\hat\mu$ the magnetic vector 
and $\mu_N$ the nuclear magneton. 
$g(I)$ can be written as a sum of proton and neutron contributions 
$g_\pi(I) + g_\nu(I)$, with 
\begin{eqnarray}
g_\tau(I)& = & \frac{1}{\mu_N \sqrt{I(I+1)}} 
\nonumber \\ && \times \left(
     g^{\tau}_l \bra \psi^I||\hat j^\tau ||\psi^I \ket 
     + (g^{\tau}_s - g^{\tau}_l) \bra\psi^I||\hat s^\tau||\psi^I\ket \right),
\label{g-form}
\end{eqnarray}
where $\tau= \pi$ or $\nu$, 
and $g_l$ and $g_s$ are the orbit and spin
gyromagnetic ratios, respectively.
We use the free values for $g_l$ 
and the free values damped by the usual 0.75 factor for $g_s$ 
\cite{BM75} 
\begin{eqnarray}
g_l^\pi &=& 1  \;\;\;\;\;\;\; g_s^\pi = 5.586 \times 0.75 
\nonumber \\
g_l^\nu &=& 0  \;\;\;\;\;\;\; g_s^\nu = -3.826 \times 0.75 .
\label{g-const}
\end{eqnarray}
We emphasize \cite{SE94} that
the g-factor in the PSM is computed directly from 
the many-body wave function 
without a semiclassical separation of the collective 
and single-particle parts. 
In particular, because of the large configuration space employed,  
there is no need to introduce a core
contribution, $g_R$, which is a model-dependent concept
and not a measurable quantity.

The calculated $Q_t(I)$ values are plotted in Fig. 3(a) as squares. 
Our calculation
gives a constant value of 9.9 $e$b up to spin $I=20\hbar$. A smooth increase
follows until $Q_t$ reaches the maximum value of 10.4 $e$b. These values
are compared with the average value ($>$ 9.5 $e$b) from experiment \cite{Exp}.  
The increase in $Q_t$ (about 12$\%$ in the spin range roughly from 
$I=20\hbar$ to $I=50\hbar$) is associated with increased contribution of
sharply down-sloping $i_{13/2}$ orbitals.  
We note that 
a larger basis deformation $\epsilon_2$ used for construction of the
shell model space will in general give rise to a larger quadrupole
moment calculated from the obtained wave functions. 
Here, what we believe is more important is the clear increase in
$Q_t$ near the band crossing region. 
Usually, a band crossing will cause
a drop in $Q_t$ values. 

To see that the major influence discussed above is mainly  
from the protons, not from the neutrons, we may study the rotational
behavior of the g-factor. Due to the opposite sign of neutron and proton 
$g_s$ in (\ref{g-const}),
effects originating 
from neutrons and protons should be easily distinguished.    
As shown by squares in Fig. 3(b), our calculations indicate a 
significant increase of the g-factor around the band crossing region. 
The increase is rather pronounced ($\sim 25\%$). 
This indicates that starting from $I=20\hbar$, components of 
the $i_{13/2}$ proton wave function
become more and more dominant. 
In contrast, no increase can be seen if the  proton 2-qp states are excluded
from the calculation, as shown by the circles in Fig. 3(b).  
The slight decrease at higher spins is because the neutron
2-qp states will
first cross the ground band (at $I=46\hbar$) in the absence of the protons. 

Thus, measurement of g-factors for several spin states and additional 
measurement of $Q_t$ should be able to test our predictions 
and provide a deeper
understanding of the
measured band. Excited bands with very extended shapes
in $^{108}$Cd might also been observed; our calculations show  
that these are likely to be 
mainly of neutron $i_{13/2}$ 2-qp structure (see Fig. 2). 

In summary, 
using the projected shell model, which has been successfully applied to a
microscopic description of nuclear superdeformed structure in several 
mass regions, we have studied the recently measured highly 
deformed rotational band 
in $^{108}$Cd. Our suggested spin values for the experimental levels 
are within 2 spin units of the experimental estimation. 
The observed irregularity in the dynamic moment of inertia is attributed
to the quasiparticle alignment of $i_{13/2}$ orbitals. 
Moreover, our calculations suggest that the 
important $i_{13/2}$ configurations
are not built on neutrons but on protons, and that the measured states have
a proton $i_{13/2}$ 2-qp structure. Rotational behavior of $Q_t$ and
the g-factor has also been studied, 
with several predictions awaiting future experimental
confirmation.   

The present work provides a theoretical framework for a detailed
structure study of nuclei 
with very extended shapes.
One may expect that rotational bands in the neighboring nuclei could 
have similar properties and may be accessible in future experiment. Work on
calculating neighboring nuclei is in progress.    
The present contribution has shown another example that the angular momentum
projection is indeed a very efficient truncation scheme for  
a shell model calculation for medium or heavy nuclei with large 
quadrupole and pairing collectivities.  

                            Research at the University of
                            Tennessee is supported by the U.~S. Department
                            of Energy through Contract No.\
                            DE--FG05--96ER40983.

\baselineskip = 14pt
\bibliographystyle{unsrt}

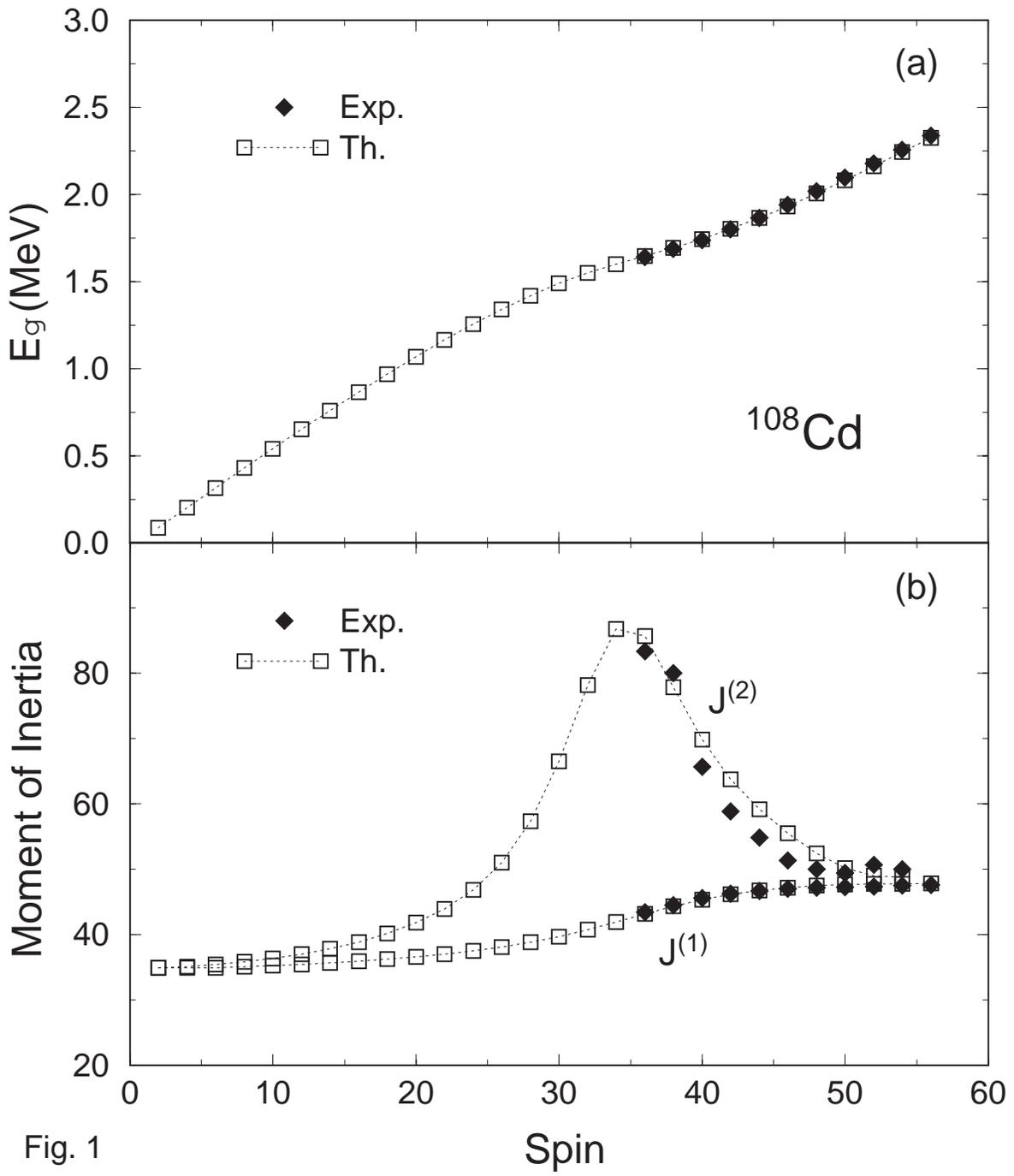
\begin{figure}
\caption{
(a) The calculated $E_\gamma (I) = E(I) - E(I-2)$ (MeV) with 
the measured $\gamma$-ray energies 
\protect\cite{Exp} best fitted to the theory curve. 
(b) Comparison of the data \protect\cite{Exp} with 
the calculated static moment of inertia 
$\Im^{(1)}=(2I-1)/E_\gamma(I)$ $(\hbar^2 \mbox{MeV}^{-1})$ and
dynamic moment of inertia $\Im^{(2)}=4/[E_\gamma(I+2)-E_\gamma(I)]$ 
$(\hbar^2 \mbox{MeV}^{-1})$. 
}
\label{figure.1}
\end{figure}

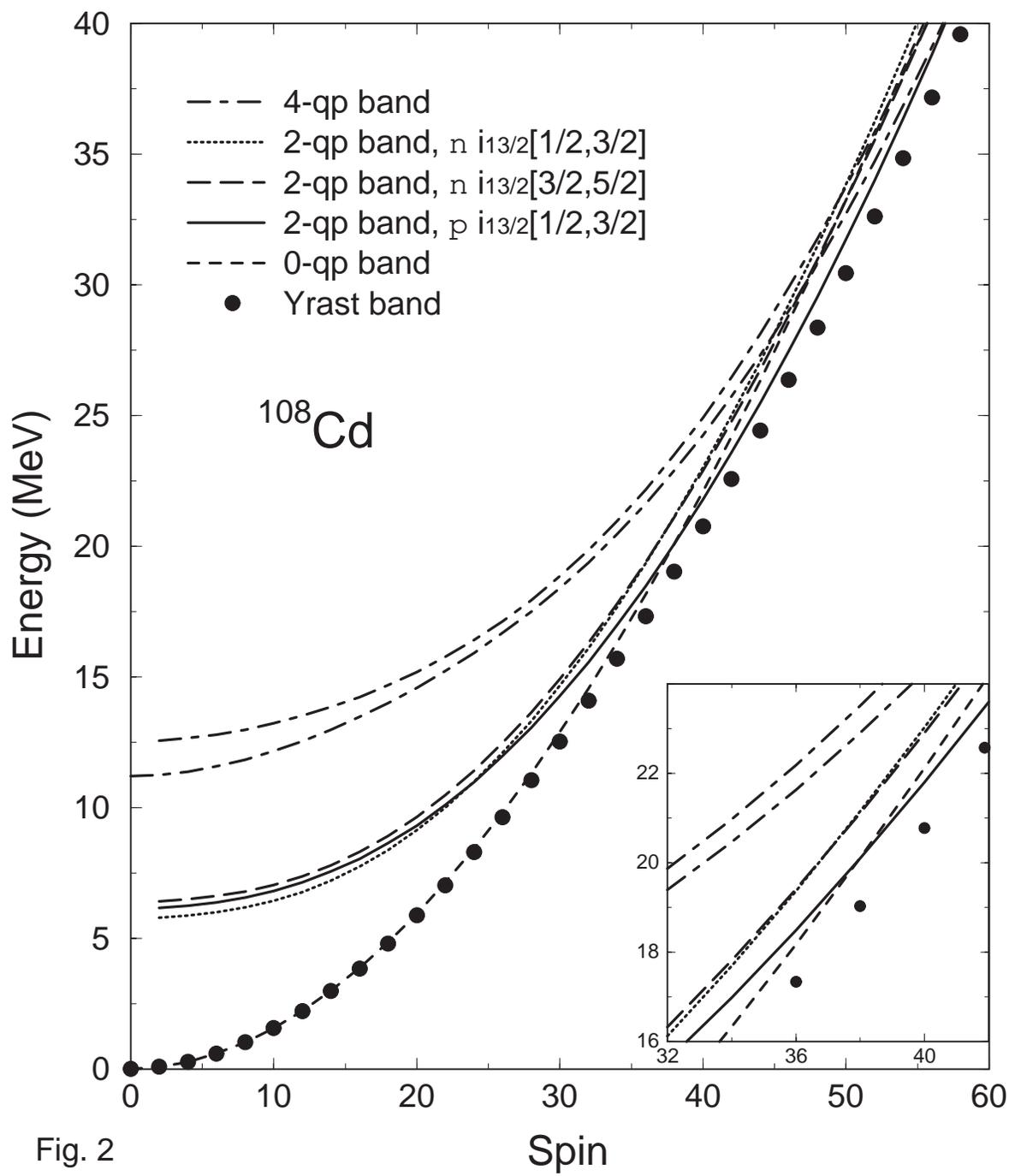
\begin{figure}
\caption{
A band diagram displaying rotational bands calculated for various
multi-qp configurations. Dots are the Yrast band energies
obtained after configuration mixing. Theoretical results presented in
Fig. 1 are computed from these energies.  
The inset is an enhancement of the spin region 
around the first band crossing
($I=32\hbar - 42\hbar$). 
}
\label{figure.2}
\end{figure}

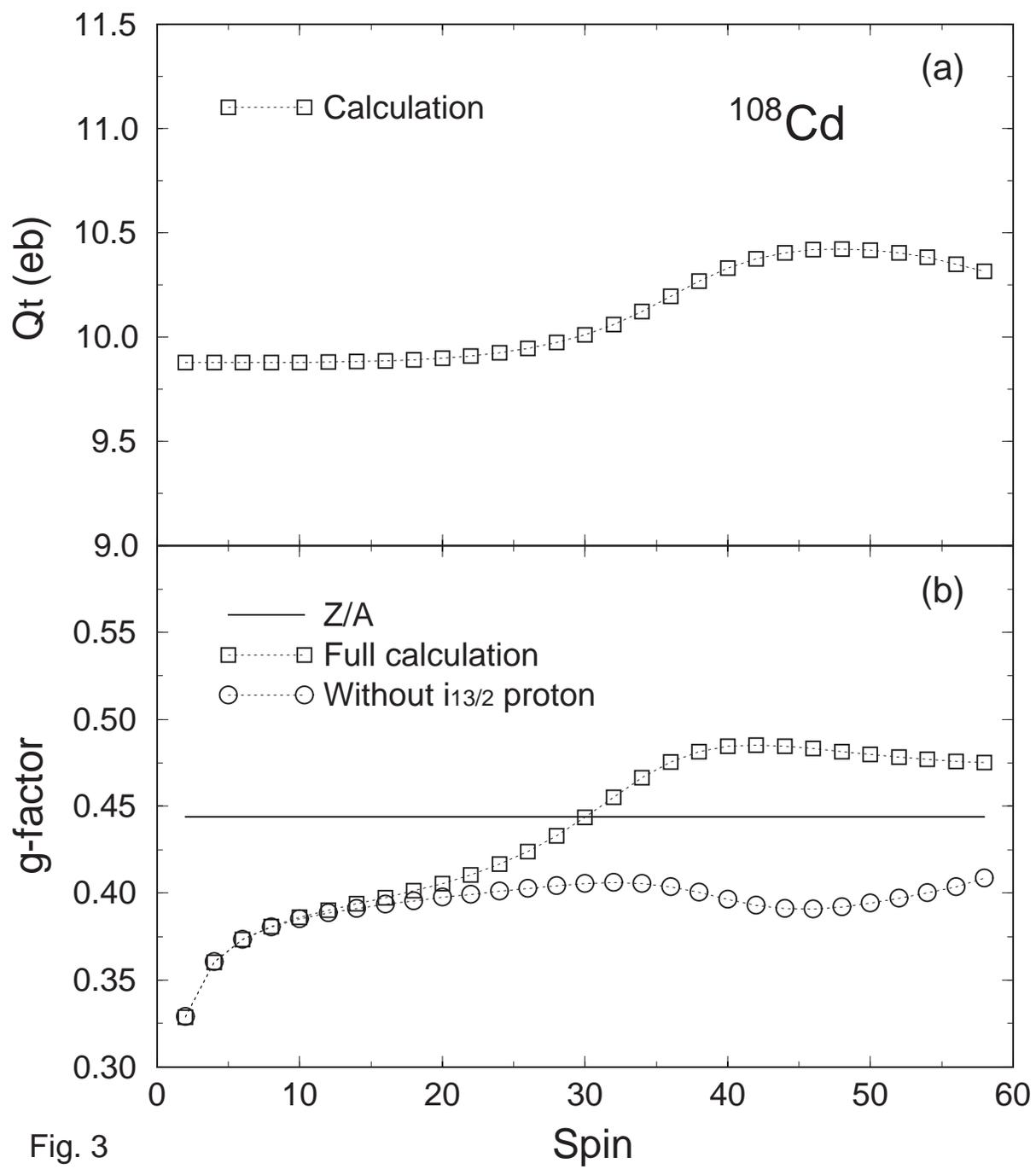
\begin{figure}
\caption{
(a) The calculated transition quadrupole moments $Q_t (e\mbox{b})$. The
experimental lower limit for the average $Q_t$ is 9.5 $e$b
\protect\cite{Exp}. 
(b) The calculated g-factors. For comparison, the rotor value Z/A is
also shown. 
}
\label{figure.3}
\end{figure}

\end{document}